\documentclass[12pt]{elsarticle}
\usepackage{hyperref}
\usepackage{graphicx}
\usepackage{dcolumn}
\usepackage{bm}
\usepackage{xcolor}
\usepackage{version}
\usepackage{amsfonts}
\usepackage{soul}
\usepackage{float}
\usepackage{mathtools}

\begin{document}

\title{Uncovering Magnetic Phases with Synthetic Data and Physics-Informed Training}

\author{A. Medina}
\address{Facultad de Ciencias Exactas, Universidad Nacional de La Plata, La Plata, Argentina.}
\address{IFLP - CIC, Departamento de Física, Universidad Nacional de La Plata, Diagonal 113 entre 63 y 64, 1900 La Plata, Argentina.}

 \author{M. Arlego}
 \address{Institute for Theoretical Physics, Technical University Braunschweig, Braunschweig, Germany}
 \address{IFLP - CONICET, Departamento de Física, Universidad Nacional de La Plata, Diagonal 113 entre 63 y 64, 1900 La Plata, Argentina.}
 \address{Facultad de Ciencias Exactas, Universidad Nacional del Centro de la Provincia de Buenos Aires, Tandil, Argentina.}

\author{ C. A.\ Lamas}
\address{IFLP - CONICET, Departamento de Física, Universidad Nacional de La Plata, Diagonal 113 entre 63 y 64, 1900 La Plata, Argentina.}

\address{Facultad de Ciencias Exactas, Universidad Nacional de La Plata, La Plata, Argentina.}

\begin{abstract}

We investigate the efficient learning of magnetic phases using artificial neural networks trained on synthetic data, combining computational simplicity with physics-informed strategies. Focusing on the diluted Ising model, which lacks an exact analytical solution, we explore two complementary approaches: a supervised classification using simple dense neural networks, and an unsupervised detection of phase transitions using convolutional autoencoders trained solely on idealized spin configurations.

To enhance model performance, we incorporate two key forms of physics-informed guidance. First, we exploit architectural biases which preferentially amplify features related to symmetry breaking. Second, we include training configurations that explicitly break $\mathbb{Z}_2$ symmetry, reinforcing the network's ability to detect ordered phases. These mechanisms, acting in tandem, increase the network’s sensitivity to phase structure even in the absence of explicit labels. We validate the machine learning predictions through comparison with direct numerical estimates of critical temperatures and percolation thresholds.

Our results show that synthetic, structured, and computationally efficient training schemes can reveal physically meaningful phase boundaries, even in complex systems. This framework offers a low-cost and robust alternative to conventional methods, with potential applications in broader condensed matter and statistical physics contexts.

\end{abstract}

\maketitle

\section{Introduction}

In recent years, machine learning (ML) techniques have found compelling applications in the study of phase transitions~\cite{Rem2019,carrasquilla2017machine,Wang2016,van2017learning,corte2021exploring,ponte2017kernel,wang2016discovering,wang2017machine,gomez2024unsupervised,mendes2021unsupervised,andreas2025,wetzel2017unsupervised,ng2023unsupervised,marashli2025identifying,jang2024unsupervised,hu2017discovering}.

The application of artificial intelligence (AI) models in physics is expanding rapidly, yet training these models on large datasets presents several challenges. One major limitation is the scarcity of extensive real-world datasets, which are often difficult and costly to obtain through experimental means. Additionally, training on large datasets requires substantial computational power, leading to high energy consumption and environmental concerns~\cite{strubell-etal-2019-energy}.

Synthetic data has emerged as a promising solution to these challenges. By generating statistically meaningful datasets that mimic real configurations, it is possible to train robust models while reducing dependency on scarce or costly data sources. Synthetic datasets also offer advantages in terms of privacy, reproducibility, and scalability~\cite{Goyal-sinthetic}. 

However, the use of synthetic data must be carefully designed. Recent work has shown that over-reliance on model-generated data, especially in generative tasks, can lead to model collapse and degraded performance~\cite{Shumailov2024}.

In this context, Acevedo et al.~\cite{Acevedo2022} studied the neural network flow of spin configurations in the two-dimensional Ising model. This flow is generated by successive reconstructions of spin configurations obtained by an autoencoder. The authors demonstrated that recursive training of the autoencoder leads to a fixed point, where successive reconstructions converge. By analyzing the reconstruction error at each step, they linked this convergence process to the intrinsic dimension of the data \cite{camastra2016intrinsic}.

Despite the aforementioned challenges associated with generative systems, we show that synthetic data can be highly beneficial for classification tasks. In a previous study~\cite{Pavioni2023}, we investigated the supervised identification of magnetic phases using dense neural networks trained on a minimal catalog of ideal synthetic configurations. By leveraging the generalization capability of a network trained on this reduced dataset, we successfully reconstructed the phase diagram of the diluted Ising model through transfer learning \cite{Zhuang-transfer}. This was achieved by classifying Montecarlo configurations across temperature and dilution, despite these variables being entirely absent from the training data.

In the present work, we extend this approach in two directions. First, we refine the supervised method by evaluating the classification performance of progressively simpler neural network architectures, aiming for computational efficiency. Second, we introduce an unsupervised method based on convolutional autoencoders trained solely on ideal configurations. This method leverages anomaly detection: configurations outside the training regime (e.g., at finite temperature or dilution) yield higher reconstruction errors, allowing for phase separation without labeled data.

To improve model performance and generalization, physics-informed machine learning (PIML) has emerged as a promising approach for embedding prior physical knowledge—such as symmetries or conservation laws—into ML models~\cite{karniadakis2021physics}. By integrating these principles directly into the architecture or training data, PIML can enhance learning efficiency and reduce the amount of data required. In our approach, we incorporate physical insight in two complementary ways: first, by introducing architectural biases by means of  ReLU activation functions, which tend to amplify features associated with symmetry breaking; and second, by including training configurations that explicitly break $\mathbb{Z}_2$ symmetry, thereby strengthening the network’s sensitivity to the emergence of ordered phases.

The remainder of the paper is structured as follows. In Section~\ref{sec:dense}, we examine the performance of simple dense neural networks for classifying magnetic phases. Section~\ref{sec:autoencoder} introduces the unsupervised classification strategy using convolutional autoencoders trained on synthetic data. Finally, Section~\ref{sec:conclusions} presents our main findings and conclusions.

\section{Training with synthetic data in simple dense neural networks}
\label{sec:dense}

In this section, we describe the creation of a synthetic data catalog and its use in training simplified Dense Neural Networks (DNN). Our goal is to demonstrate that synthetic data enables fast, efficient training while achieving strong performance in magnetic state classification. We also highlight the generalization capabilities of the resulting models.

Traditional training data for phase classification often come from experimental measurements or numerical simulations, which can be time-consuming and require prior knowledge of the system's phases.
Here, we adopt a recent approach that builds a synthetic training catalog composed of hand-crafted spin configurations representing ideal magnetic phases. Pavioni et al.\cite{Pavioni2023} showed that such catalogs allow neural networks to generalize and effectively classify more complex scenarios. In this work, we revisit the (site) diluted Ising model on a two-dimensional lattice, a paradigmatic system in condensed matter physics, to further explore this approach.

The model consists of $N \times N$ sites, each with a spin variable $S_i = \pm 1$, defining the global configuration $\{S_i\}$. Defects are introduced by randomly removing spins, and the system's Hamiltonian is given by
\begin{equation}
        \mathcal{H}=-\sum_{ i,j } ^{N} J_{i,j}\epsilon_{i} \epsilon_{j} S_{i}S_{j},
        \label{H}
\end{equation}
where $\epsilon_i = 1$ indicates an occupied site and $\epsilon_i = 0$ an empty one. We focus on ferromagnetic nearest-neighbor couplings on a square lattice $(J_{i,j}=J>0)$.

This diluted model mimics materials with impurities, where the spin lattice density is $\rho = \sum_{i}^{N^2} \epsilon_i / N^2$. The pure case $(\rho = 1)$ exhibits a second-order phase transition from a high-temperature paramagnetic state to a low-temperature ferromagnetic state, characterized by long-range correlations and critical behavior.
As $\rho$ decreases, the critical temperature $T_c$ also drops. Below the percolation threshold $\rho_c$, the system remains disordered at all temperatures \cite{stauffer2018introduction}. For the square lattice, percolation density was precisely estimated as $\rho_c \simeq 0.59274621(13)$ \cite{Newman2000}.

\subsection{Dense Neural Networks}

\begin{figure}[h]
    \centering
    \includegraphics[width=0.5\textwidth]{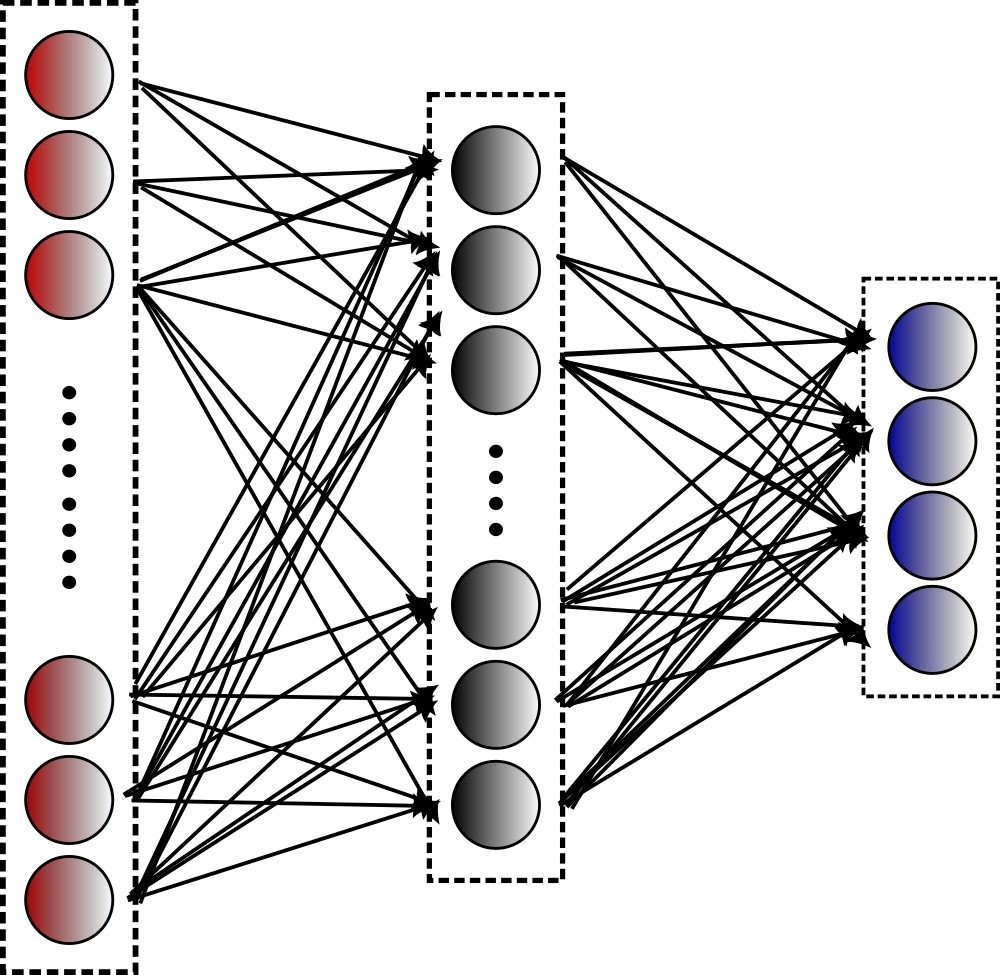}
    \caption{Schematic representation of the DNN used in this study. The input layer has a fixed size of 1600 neurons ($40\times 40$ square spin lattice). The hidden layer size was varied to analyze performance with a reduced number of neurons. The output layer size corresponds to the number of magnetic phases used during training.}
    \label{fig:myNN}
\end{figure}

To evaluate DNN performance, we begin by predicting the probability of each magnetic phase. The input layer has 1600 neurons, corresponding to a $40 \times 40$ spin lattice. The output layer matches the number of magnetic phases in the catalog: ferromagnetic (F), Néel (N), stripes (S), and random (P). The first three represent $T=0$ ordered phases, and the last corresponds to the $T=\infty$ paramagnetic state. No finite-temperature configurations are included in the training data. The network structure is shown in Fig.~\ref{fig:myNN}.

To keep the architecture simple, we used a single hidden layer. Its size is the only adjustable parameter once the number of magnetic classes is fixed.

Since the fundamental principles of neural networks and their training procedures are well established, we focus here on the specific architecture employed and the generation of training data in our study. For further background, including both theoretical and practical aspects, see Refs.~\cite{bengio2017deep,géron2022hands,chollet2021deep}.

After training on synthetic data, the DNN was used to classify spin configurations obtained via Montecarlo simulations of the Hamiltonian~(\ref{H}). For training details, see Ref.~\cite{Pavioni2023}. At each temperature, classification results were averaged over 300 configurations to estimate class probabilities.

\begin{figure}[ht]
    \centering
    \includegraphics[width=0.8\linewidth]{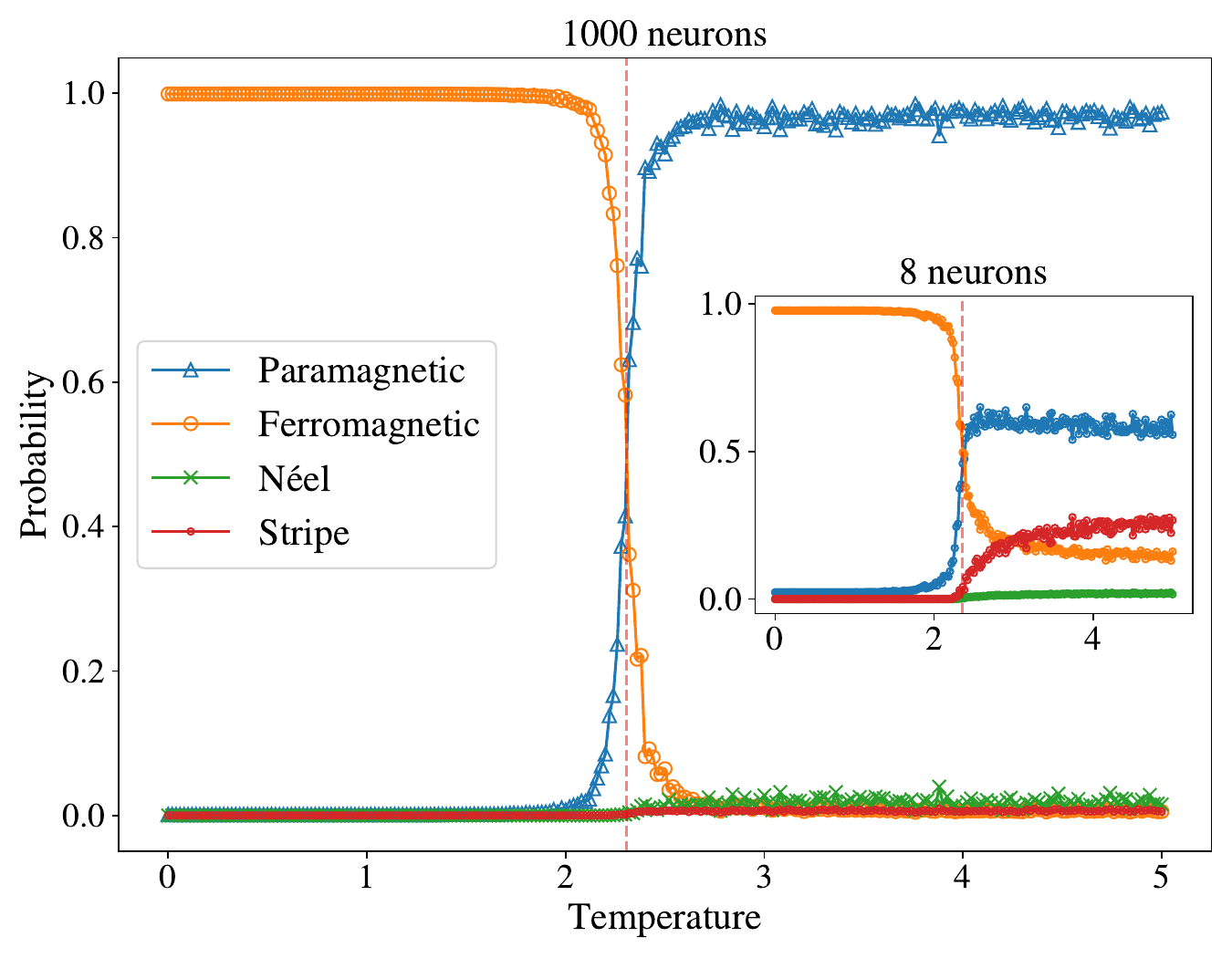}
    \caption{Probability distribution of each configuration as a function of temperature for the pure Ising model on a $40 \times 40$ square spin lattice, computed using a DNN with a hidden layer of 1000 neurons (main panel) and 8 neurons (inset).}
    \label{fig:probabilities}
\end{figure}

Figure~\ref{fig:probabilities} shows these probabilities for the pure Ising model, using 1000 and 8 hidden neurons (inset). The model identifies two dominant phases—ferromagnetic and paramagnetic—within the range $T/J = 0$ to $T/J = 5$, effectively placing the system in either an ordered (F) or disordered (P) state.

The dominant class probability determines the predicted phase at each temperature. At high temperatures, disordered (P) features dominate; at low temperatures, the ferromagnetic (F) phase prevails. In between, probability trends shift, indicating a transition.

To estimate the transition temperature, we applied a “maximum confusion” criterion: if class probabilities are comparable—particularly between F and P—we interpret this as the system being near a phase transition. The threshold $p[\text{F}] > 0.5$ is used to identify predominantly ferromagnetic configurations.

\subsection{Effect of Hidden Neurons on Network Accuracy}

We analyze how the number of hidden neurons affects the DNN’s ability to classify magnetic phases. As seen in Fig.~\ref{fig:probabilities}, this parameter has a strong influence on model performance.
A larger hidden layer improves identification of the transition temperature but increases computational cost and the risk of overfitting. Too few neurons, on the other hand, hinder accurate phase discrimination.

\begin{figure}[h]
    \centering
    \includegraphics[width=\linewidth]{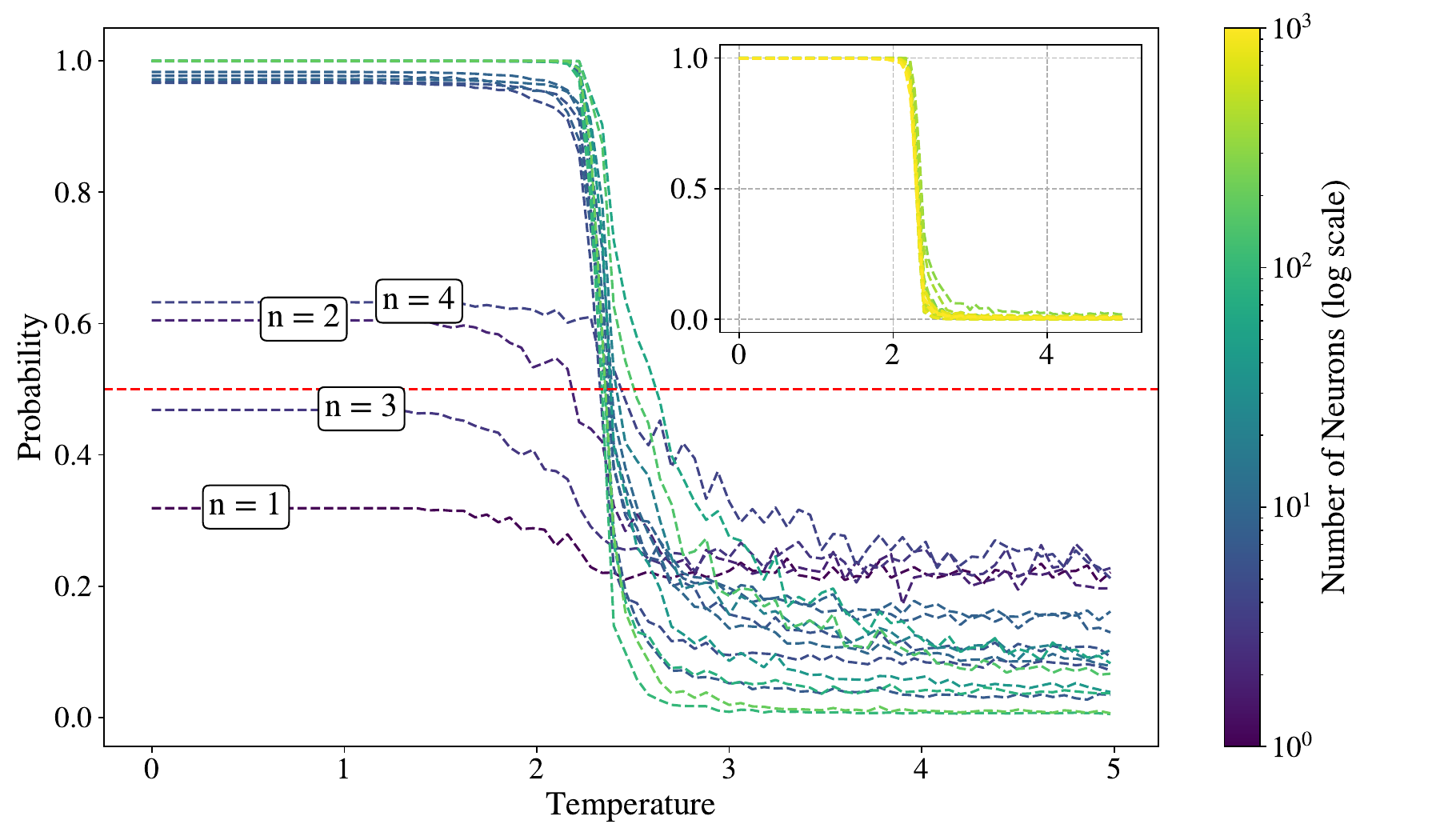}
    \caption{Probability distribution of each configuration as a function of temperature for the pure Ising model on a $40 \times 40$ square spin lattice, computed using a DNN with a hidden layer containing up to 400 neurons (main panel) and between 400 and 1000 neurons (inset). Note the marked improvement in performance when using 5 or more hidden neurons (main panel).}
    \label{fig:ferro}
\end{figure}

To determine a suitable size for the hidden layer, we trained multiple models with varying numbers of neurons using synthetic data and evaluated their outputs. Figure~\ref{fig:ferro} shows that small hidden layers yield high variability (main panel), while larger ones produce sharper transitions resembling step functions (inset). Although this suggests better performance, it also implies a greater number of free parameters.

For $n = 2$, the probability of a ferromagnetic configuration first exceeds 0.5; with $n = 4$, it improves slightly. For $n \geq 5$, the model rapidly reaches near-perfect classification of the ferromagnetic phase.

\begin{figure}[h]
    \centering
    \includegraphics[width=\linewidth]{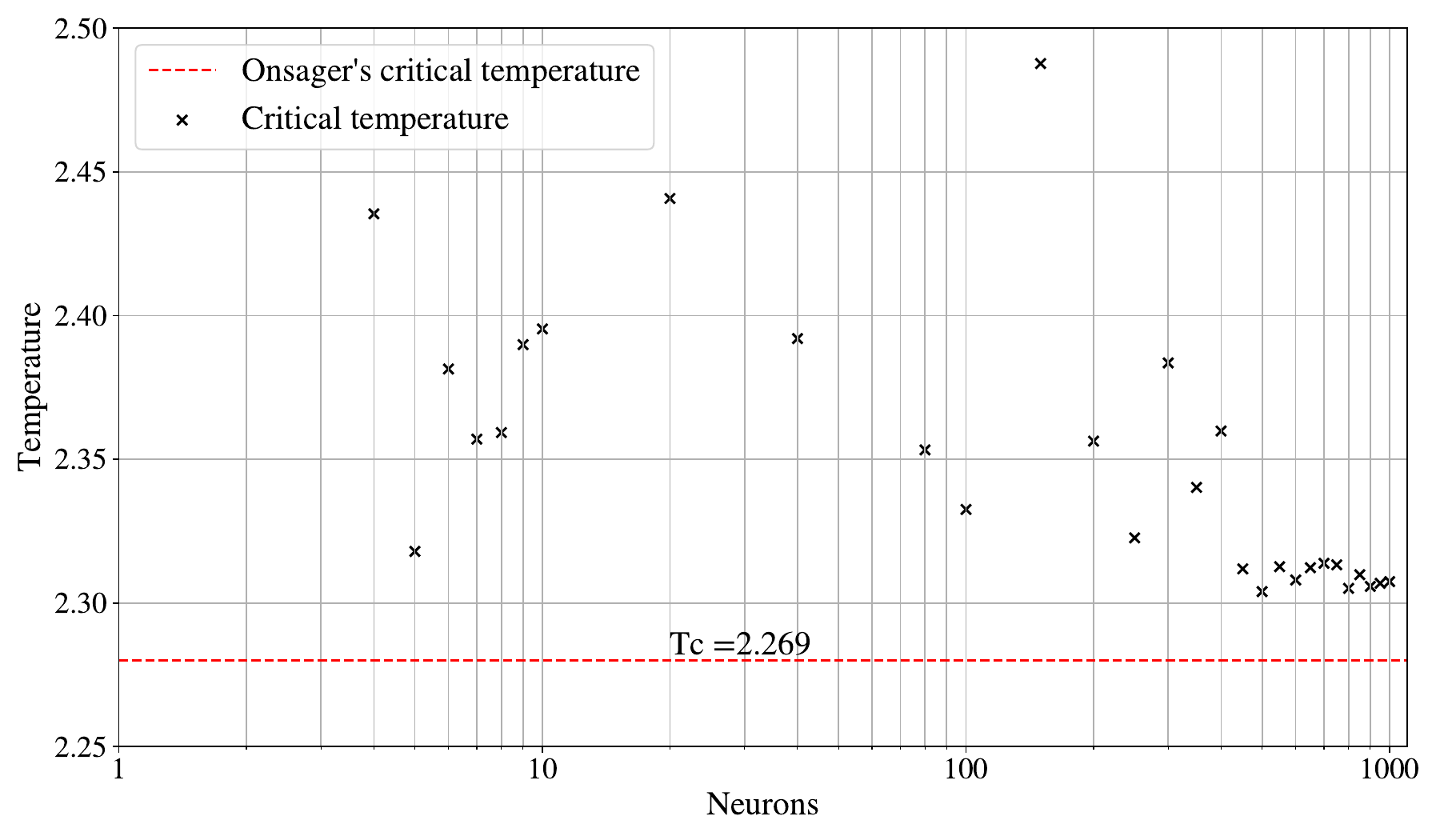}
    \caption{Critical temperature predicted by the DNN as a function of the number of neurons in the hidden layer, for the pure Ising model on a $40 \times 40$ square spin lattice. Although a marked improvement in performance is observed for $n \geq 5$ (see Fig. \ref{fig:ferro}), a stable prediction of the critical temperature is not achieved until the number of hidden neurons exceeds 400. The red dashed line indicates the exact critical temperature in the thermodynamic limit.}
    \label{fig:critical_temperature}
\end{figure}

We then studied the predicted transition temperature using the maximum confusion criterion. Figure~\ref{fig:critical_temperature} shows that estimates are consistently above the theoretical value due to finite-size effects (it should be noted that the Montecarlo simulations were performed for discrete values of the temperature with step $\Delta T= 0.02$).
Finite-size scaling, not pursued here, would be required to correct this bias.

As shown in Fig.~\ref{fig:critical_temperature}, the predicted $T_c$ fluctuates for $n$ between 10 and 300, stabilizing only when $n > 400$, closer to the expected critical temperature.

\subsection{Transition temperature estimation in the dilute Ising model}

We now apply the trained neural network to classify phases in the diluted Ising model, using only idealized configurations from a fully occupied lattice ($\rho = 1$). No finite-temperature or vacancy information was included during training, highlighting the model’s generalization ability.

To estimate the transition temperature for various spin densities $\rho$, we trained a DNN with two output neurons (ferromagnetic and paramagnetic). At low $\rho$, the presence of Néel and stripe configurations introduces ambiguity, and $\rho = 0.6$ is the first case where the ferromagnetic probability exceeds 50\%.

\begin{figure}[t]
    \centering
    \includegraphics[width=\linewidth]{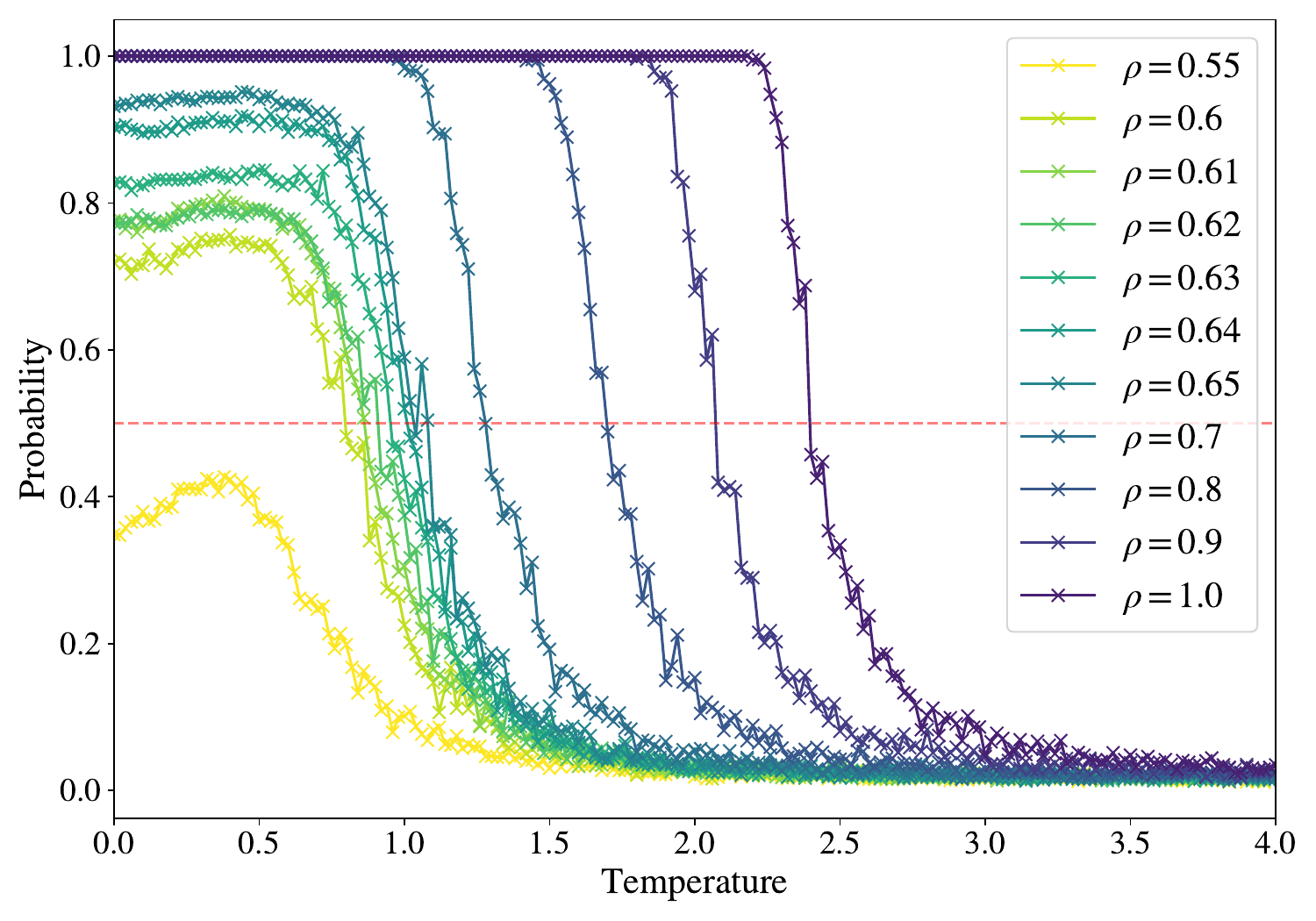}
    \caption{Probability of ferromagnetic configurations as a function of temperature for various spin densities in the diluted Ising model. The DNN is trained with ferromagnetic and paramagnetic configurations. For $\rho = 0.55$, the DNN does not find an ordered state for any temperature and predicts $0.55<\rho_c<0.6$.}
    \label{fig:ferro_predictions}
\end{figure}

Figure~\ref{fig:ferro_predictions} shows the ferromagnetic probability versus temperature for different $\rho$ values. As $\rho$ decreases, the probability drops, falling below 0.5 for $\rho < 0.6$, indicating the system remains disordered at all temperatures.

\begin{figure}[t]
    \centering
    \includegraphics[width=0.9\linewidth]{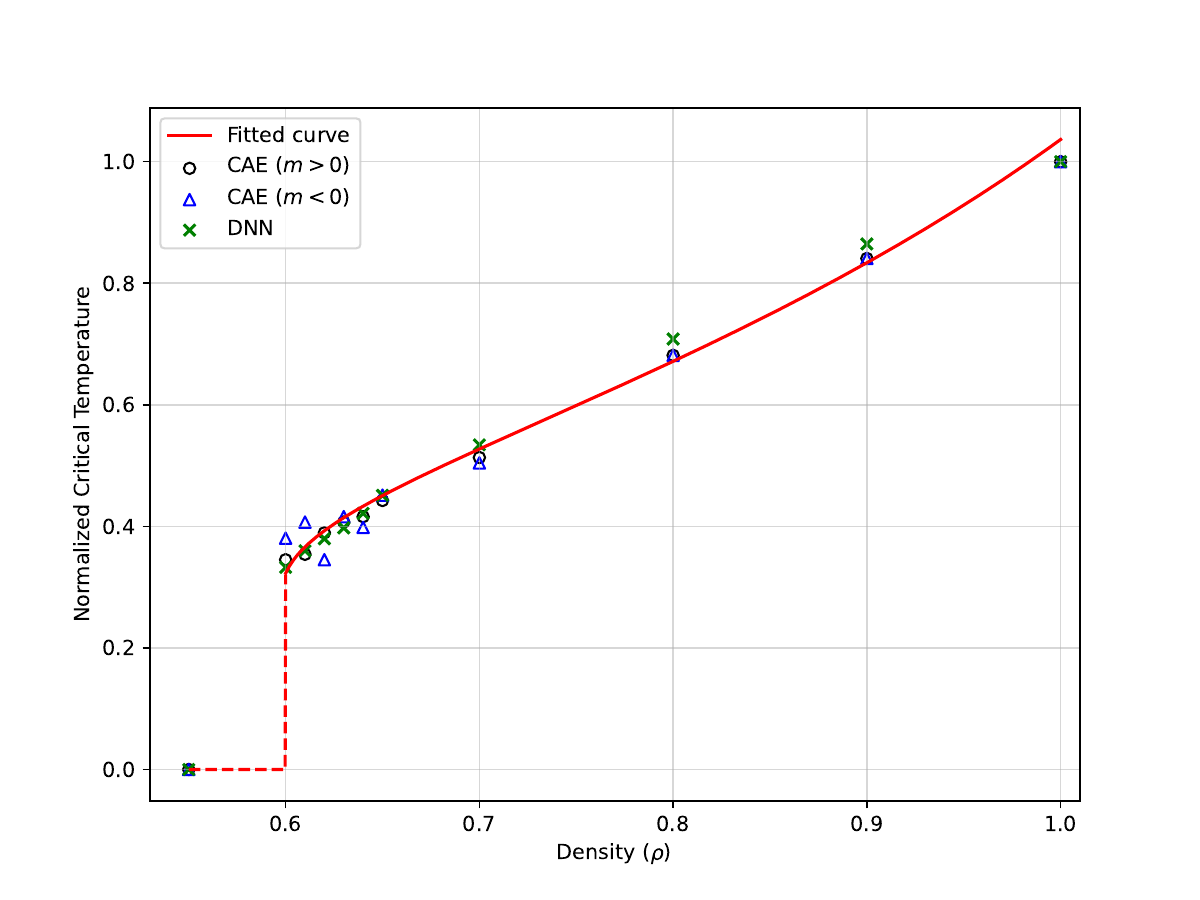}
   \caption{(Color Online) Transition temperature as a function of spin density, normalized to $T_c(\rho = 1)$. Green crosses correspond to estimates obtained from the supervised DNN. The solid red line shows a fit to the form $T_c = -\frac{K}{\ln{(\rho - \rho_c)}} + A$. Circular and triangular markers indicate estimates from unsupervised methods discussed in Section \ref{sec:autoencoder}. For $\rho = 0.55$, none of the methods employed detect a transition.
}
    \label{fig:critical_temperature_perco}
\end{figure}

The transition temperature was estimated for each $\rho$ and plotted as green crosses in Fig.~\ref{fig:critical_temperature_perco}, alongside a fit to the form \cite{Pavioni2023}
\begin{equation}
\frac{T_c(\rho)}{T_c(1)} =  - \frac{K}{\ln{(\rho - \rho_c)}}+A.
\label{eq:fit}
\end{equation}

Here, temperatures are normalized to $T_c(\rho = 1)$. Fitting yields $K=0.77(3)$, $A=0.18(2)$, and a critical density $\rho_c = 0.597(5)$.

The absence of ordering at low $\rho$ signals percolation. The estimated threshold $\rho_c \approx 0.597$ aligns well with known values in the literature\cite{Newman2000}.

Figure~\ref{fig:critical_temperature_perco} also includes estimates from two unsupervised methods—marked with circles and triangles—that will be discussed in the following section.

\section{Synthetic Anomaly Detection }
\label{sec:autoencoder}

\begin{figure}[ht]
    \centering
    \includegraphics[width=0.9\linewidth]{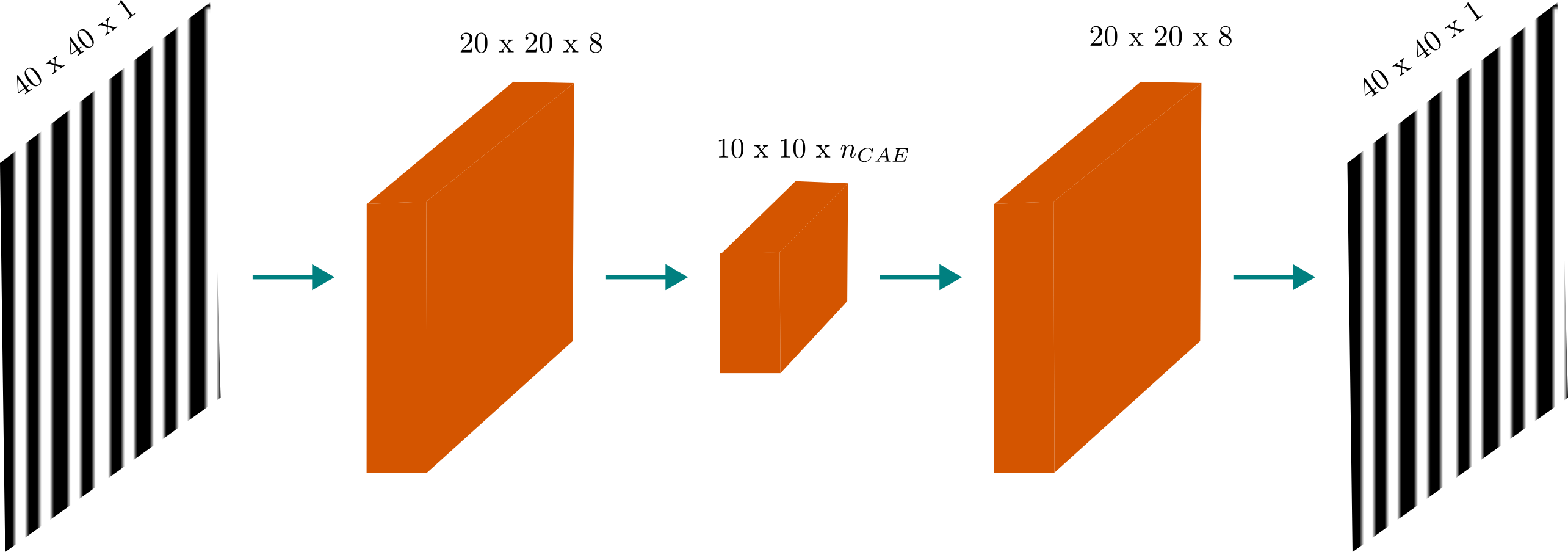}
    \caption{Pipeline of the convolutional autoencoder. The input images are $40 \times 40$ synthetic spin configurations. A first convolutional layer with $8$ filters of size $3 \times 3$ is applied, followed by an intermediate convolutional layer with $n_{\text{CAE}}$ filters of size $3 \times 3$. Finally, a deconvolutional layer with $8$ filters of size $3 \times 3$ reconstructs the output.
}
    \label{fig:table_autoencoder}
\end{figure}

In the previous section, we examined a classification model based on dense neural networks, traditionally used in supervised learning settings. However, unlike conventional supervised learning—where labeled data is typically obtained through manual annotation—we employed a synthetic dataset where both the configurations and their corresponding phase labels were generated programmatically. While this still qualifies as supervised learning in a technical sense due to the presence of labels, it departs from standard practices by removing the reliance on empirical or human-labeled data.

In this section, we shift to a fully unsupervised learning strategy to detect the magnetic phase transition. Specifically, we exploit anomaly detection techniques applied to magnetic phase classification, as explored in prior works~\cite{ng2023unsupervised,Acevedo2021}.

In standard anomaly detection applications, autoencoders are typically trained on real-world data that is representative of a given class or regime, though not necessarily labeled. Once trained, the autoencoder estimates whether new data belong to the same distribution by evaluating reconstruction error: significant deviations suggest the presence of anomalies. 

Here, we take this idea to the extreme by training a convolutional autoencoder (CAE) \cite{bengio2017deep} exclusively on synthetic data representing idealized magnetic configurations from ordered phases. As in the dense network case, the training set does not include any Montecarlo samples corresponding to finite-temperature or disordered configurations. This setup ensures that the model is exposed only to symmetry-breaking patterns during training, enabling it to identify paramagnetic states—entirely absent from the training distribution—as “anomalous” due to their higher reconstruction error.

The autoencoder architecture consists of a simple convolutional encoder – decoder with a bottleneck latent layer (see Fig.~\ref{fig:table_autoencoder}). We systematically evaluated the model’s performance across different architectural variants by changing the number of filters in the intermediate convolutional layer. The best trade-off between complexity and accuracy was found with $n_{\text{CAE}}=8$ filters, a configuration we adopt for the remainder of our experiments.

The network was trained for 150 epochs using cross-entropy loss. ReLU activations were applied in all hidden layers, while a Sigmoid function was used at the output layer \cite{chollet2021deep}. This architectural choice introduces a crucial inductive bias into the model: the ReLU activation inherently favors positive values and thus explicitly breaks $\mathbb{Z}_2$ spin-inversion symmetry. In the context of magnetic phases, this means that “up” magnetized configurations are more easily reconstructed than “down” ones, even when both are part of the training set.

This symmetry-breaking bias is not a flaw but a deliberate design feature. By amplifying differences associated with $\mathbb{Z}_2$ symmetry, the network becomes more sensitive to transitions between ordered and disordered phases. While it is theoretically possible to symmetrize the network to reduce this bias, doing so would weaken the model’s ability to detect when symmetry is restored, as occurs in the paramagnetic regime.

Therefore, instead of aiming for optimal reconstruction fidelity, we retain the asymmetry introduced by ReLU activations as a mechanism to magnify the contrast between phases. This architectural strategy enhances the CAE’s utility as a detector of symmetry restoration, linking reconstruction error to phase structure in a physics-informed way.

A similar approach to detecting symmetry breaking through anomaly detection was recently explored by Acevedo et al.~\cite{Acevedo2021}, who used variational autoencoders to identify both $\mathbb{Z}_2$ and $\mathbb{Z}_4$ symmetry breaking. However, their models required training on large datasets obtained from Montecarlo simulations, making the process computationally demanding. In contrast, our results show that a simple and efficient training strategy based solely on synthetic data is sufficient to reveal symmetry breaking and characterize magnetic phase transitions.

\subsection{Pure Ising Model: Symmetry Bias and Reconstruction Error}

\begin{figure}[ht]
   \centering
    \includegraphics[width=0.99\linewidth]{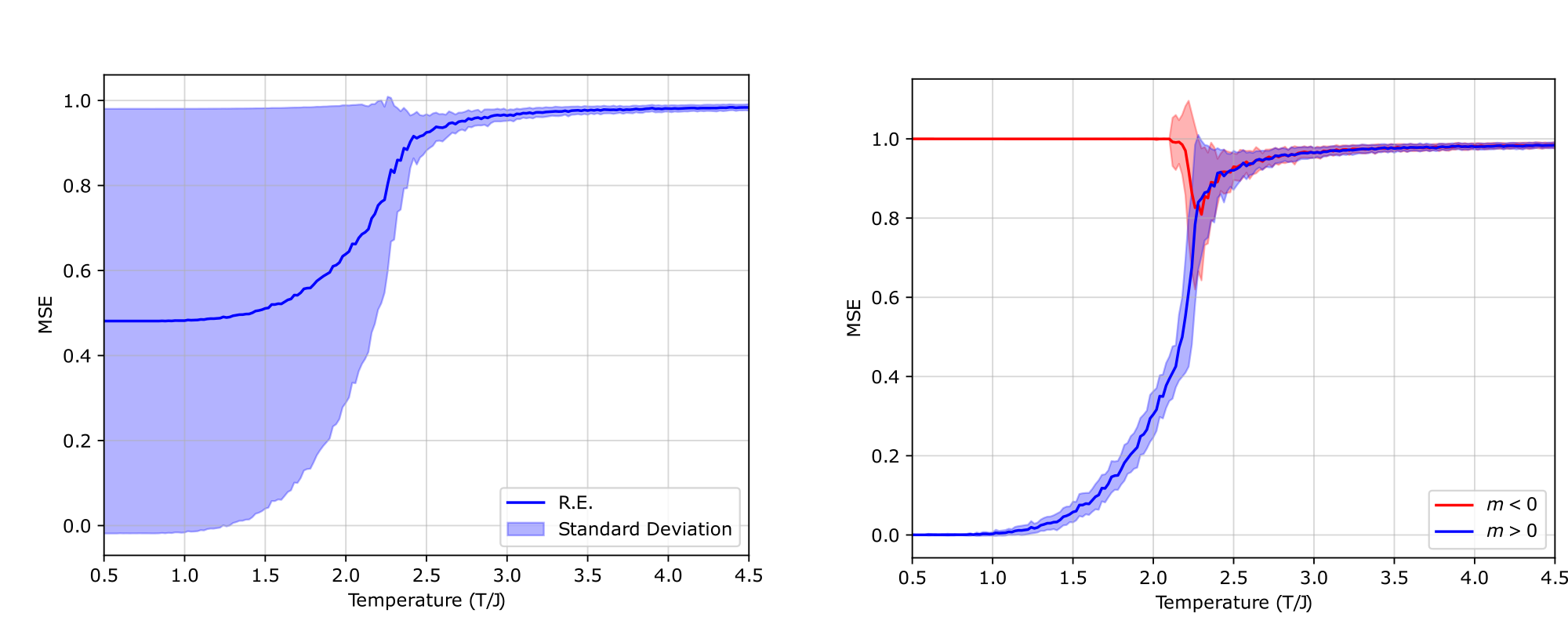}
    \caption{ Average Mean square Error as a function of T. Left panel: The blue line represent the average over 300 configurations. Light-blue area corresponds to the SD.  Right panel: Average Means square Error and SD corresponding to positive (blue) and negative (red) total magnetization states vs. T.}
    \label{fig:errorMean}
\end{figure}

While it is expected that the autoencoder successfully reconstructs the simple, idealized configurations on which it was trained (see Appendix), it is particularly notable that it can generalize this reconstruction ability to more complex, unseen configurations without any retraining. A visual inspection of these reconstructions suggests that the discrepancy between input and reconstructed images increases with temperature. In the following analysis, we quantify this effect using the mean squared error (MSE) between the input and the output of the autoencoder. MSE is given by the expression \cite{chollet2021deep}

\begin{equation}
    MSE = \frac{1}{N^2} \sum_{i, j=1}^N |x_{ij} - \hat{x}_{ij}|^2,
    \label{eq:mse}
\end{equation}
where $x_{ij}$ are the pixels in the input image while $\hat{x}_{ij}$ corresponds to the reconstructed image. Figure~\ref{fig:errorMean} shows the MSE as a function of temperature for the pure Ising model. Each point corresponds to an average over 300 independent Montecarlo configurations. The solid blue line in the left panel indicates the average MSE, while the shaded region represents the standard deviation.

As the temperature increases, the system transitions from an ordered to a disordered phase. This is reflected in the growth of the average MSE: in the high-temperature paramagnetic phase, the configurations differ substantially from those seen during training, and are therefore reconstructed poorly by the network. This behavior is captured by the solid blue line, which increases with temperature and eventually stabilizes at a high value in the disordered regime. Since disordered samples were not included in the training set, the network treats them as anomalous inputs.

A striking feature in the left panel is the large variance observed at low temperatures, where the CAE was expected to perform well. To understand this behavior, we split the configurations based on the sign of their total magnetization. The right panel of Fig.~\ref{fig:errorMean} shows the mean MSE separately for positively and negatively magnetized states. It reveals a clear split: configurations with positive magnetization ($m > 0$) are reconstructed with low error, while those with negative magnetization ($m < 0$) exhibit much higher MSE.

This splitting is a consequence of spontaneous $\mathbb{Z}_2$ symmetry breaking in the ordered phase, combined with the ReLU-induced architectural bias of the network. Although positive and negative magnetization states are physically equivalent, the ReLU activation functions favor positive outputs, resulting in better reconstruction of “up” states compared to “down” ones.

Remarkably, the MSE for $m < 0$ configurations can even exceed that of disordered high-temperature configurations. This is because negative configurations consist entirely of $-1$ values, which ReLU tends to map to zero, yielding the largest possible reconstruction error—one per site. In contrast, high-temperature configurations are composed of random mixtures of $-1$ and $1$. Although unfamiliar to the network, this randomness allows it to reconstruct some regions reasonably well by forming local patches that resemble trained patterns, resulting in a lower overall MSE than in the fully negative case (See Appendix).

\subsection{Diluted Ising Model and Percolation Threshold}

We now evaluate the generalization ability of the autoencoder in a more challenging setting: the \textit{diluted} Ising model, in which a fraction of the spins are randomly removed and replaced by non-magnetic sites. The Hamiltonian for this system is given in Eq.~(\ref{H}). It is important to emphasize that the autoencoder was trained exclusively on synthetic configurations corresponding to fully occupied lattices, with no vacancies or thermal fluctuations.

Despite the increased complexity introduced by dilution, the anomaly detection strategy remains effective in capturing the key physical features of the system. In particular, the reconstruction error continues to reflect the transition from ordered to disordered phases.

\begin{figure}[ht]
    \centering
    \includegraphics[width=0.8\linewidth]{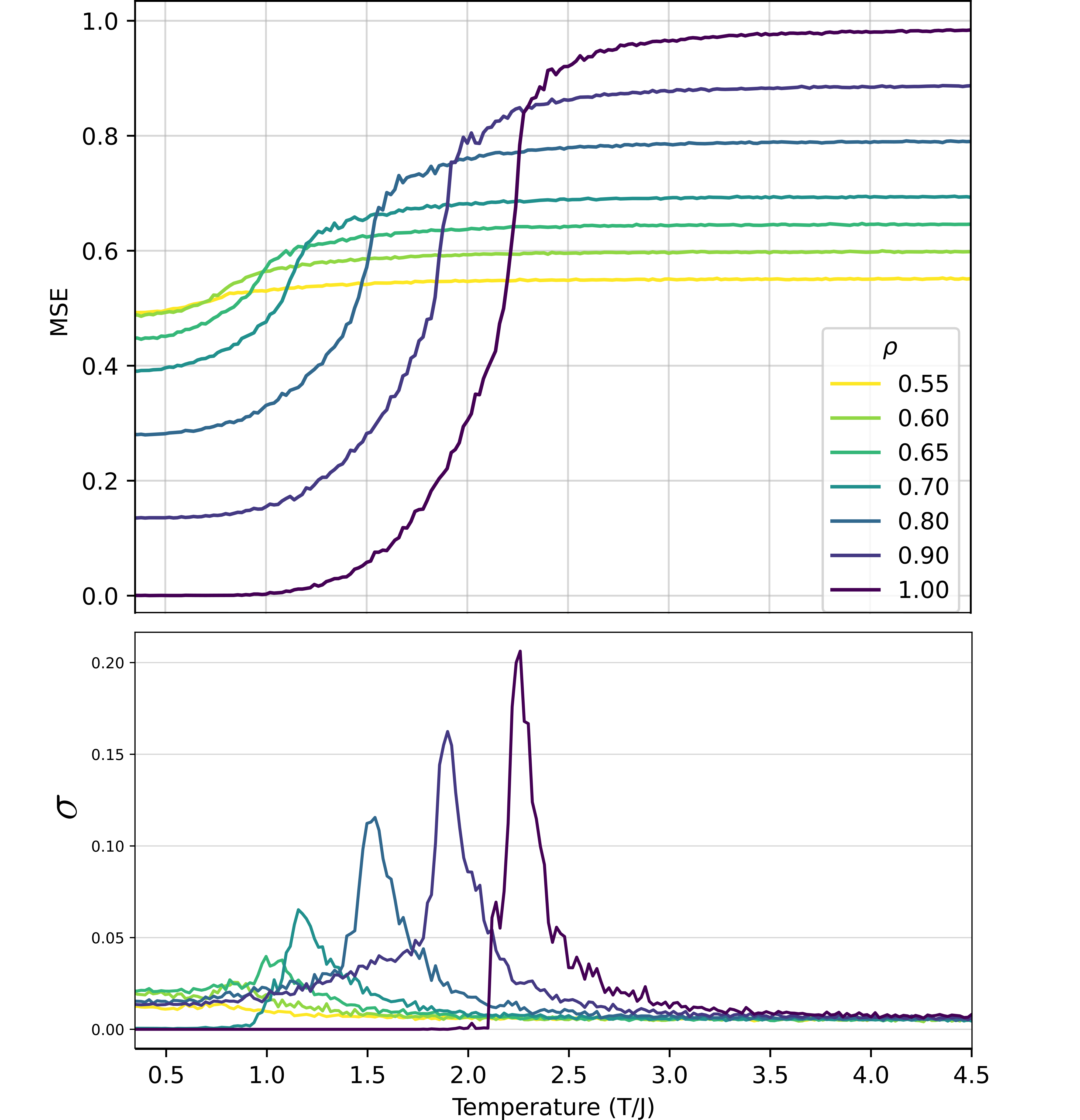}
    \caption{Top: average MSE as a function of temperature corresponding to 300 Montecarlo-generated spin patterns for several values of $\rho$. Bottom: Corresponding standard deviation}
    \label{fig:errorVsRho}
\end{figure}

Top panel in Fig.~\ref{fig:errorVsRho} shows the average MSE as a function of temperature for various spin densities $\rho$. As $\rho$ decreases, the temperature at which the MSE grows abruptly shifts to lower values. This behavior is consistent with the suppression of magnetic order. The most pronounced variation in MSE —particularly when configurations are separated by magnetization— occurs near the transition, enabling the use of the network’s output to estimate the critical temperature.

To do so, we adopt a criterion based on the \textit{maximum variance} (or standard deviation (SD)) of the reconstruction error across configurations at fixed temperature. This approach, previously employed as a proxy for criticality \cite{Pavioni2023}, is particularly effective in the high-density regime, where the phase transition remains sharp. The results obtained with this method are shown in Fig.~\ref{fig:critical_temperature_perco}, and are in good agreement with those from the supervised dense neural network and with analytical approximations for the diluted Ising model.

However, as the spin density approaches the percolation threshold, the transition becomes increasingly smooth, and the peak in variance becomes less pronounced, as observed in the bottom panel of Fig.~\ref{fig:errorVsRho}. In this low-density regime, the maximum-variance criterion becomes unreliable for locating the transition. To address this limitation, we propose a complementary strategy: we track the persistence of the MSE splitting between positively and negatively magnetized configurations at low temperatures.

This splitting, induced by the network’s ReLU bias, is typically robust in the ordered phase, with negative magnetization states showing higher reconstruction error. As $\rho$ decreases, however, the two plateaus gradually converge, reflecting the weakening of magnetic order. Eventually, the network can no longer distinguish between the two magnetization sectors, indicating the loss of long-range order and the approach to the percolation threshold.

\begin{figure}[t]
    \centering
    \includegraphics[width=\linewidth]{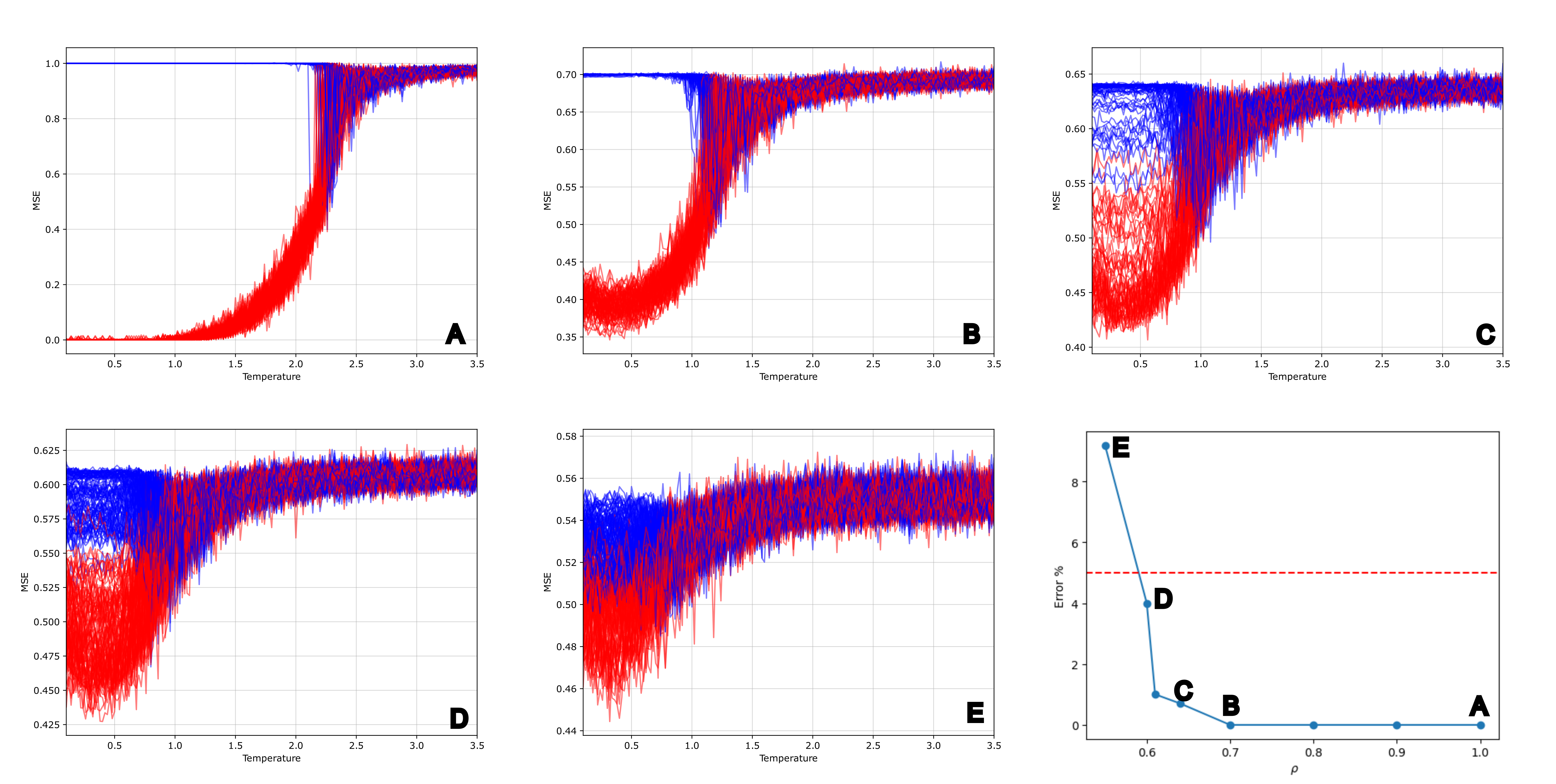}
    \caption{MSE corresponding to the reconstruction of Montecarlo configurations after training the autoencoder on synthetic ordered configurations. Panels A to F correspond to $\rho=1, 0.7, 0.64, 0.61$, and $0.55$, respectively. The bottom-right panel shows the percentage of configurations incorrectly classified at low temperatures. The red horizontal line marks the $5\%$ misclassification threshold used in this work.}
    \label{fig:SplitClosed}
\end{figure}

To quantify this effect, we analyze whether the reconstruction-error splitting remains detectable. Due to the architectural bias introduced by ReLU activations, negatively magnetized configurations typically exhibit higher reconstruction error than positive ones. As $\rho$ decreases, this difference shrinks and eventually vanishes (see Fig.~\ref{fig:SplitClosed}).

To formalize a detection criterion, we apply the K-means clustering algorithm \cite{géron2022hands} to the MSE curves at low temperature, aiming to group them into two clusters. We then assess the clustering accuracy by comparing the assigned labels to the actual sign of the total magnetization, used as ground truth. As dilution increases, the distinction between clusters weakens, and classification accuracy deteriorates.

We consider the splitting detectable as long as the misclassification rate remains below $5\%$. Using this criterion, we identify a percolation threshold around $\rho = 0.6$: for $\rho \geq 0.6$, the clustering accuracy exceeds $96\%$, whereas for $\rho = 0.55$, it falls to approximately $90\%$. 
This value allows us to complete the estimate of the transition temperature as a function of spin density in Fig. \ref{fig:critical_temperature_perco} and are consistent with previous estimates based on alternative techniques~\cite{Zhang1982,Newman2000,Newman2001}.

In summary, the convolutional autoencoder—trained exclusively on synthetic, ordered configurations—successfully detects the transition from magnetic to paramagnetic phases across a broad range of spin densities. Its ability to capture symmetry breaking, even in the presence of disorder, demonstrates the power of unsupervised anomaly detection as a tool for identifying phase transitions with minimal computational cost and no reliance on labeled data.

\subsection{Relaxed Anomaly Detection: Symmetry Bias-Complexity Interplay}

Up to this point, we have shown that a convolutional autoencoder trained exclusively on idealized low-temperature configurations—such as uniform, Néel, and stripe patterns—can successfully detect the transition to the disordered phase through an anomaly detection strategy. Disordered configurations, which were absent from training, are naturally reconstructed with higher error, serving as indicators of unfamiliar or anomalous input.

However, this setup assumes a strict anomaly detection regime, where anomalous samples lie entirely outside the training distribution. To extend this framework and explore the network’s internal dynamics in a more nuanced setting, we relaxed this condition by including fully disordered configurations (i.e., random $\pm 1$ spin patterns) in the training set. While this no longer qualifies as pure anomaly detection, it allows us to probe how the autoencoder responds to patterns with varying intrinsic complexity and symmetry properties across the full phase diagram.

This modification introduces a richer interplay between two key factors: the structural complexity of configurations and the symmetry biases encoded in the network architecture. Ordered phases, which break $\mathbb{Z}_2$ symmetry, possess low intrinsic dimensionality and are easier to reconstruct in an unbiased context, while disordered paramagnetic configurations—though now part of the training set—remain more difficult due to their higher variability and intrinsic degrees of freedom.

\begin{figure}[t]
    \centering
    \includegraphics[width=\linewidth]{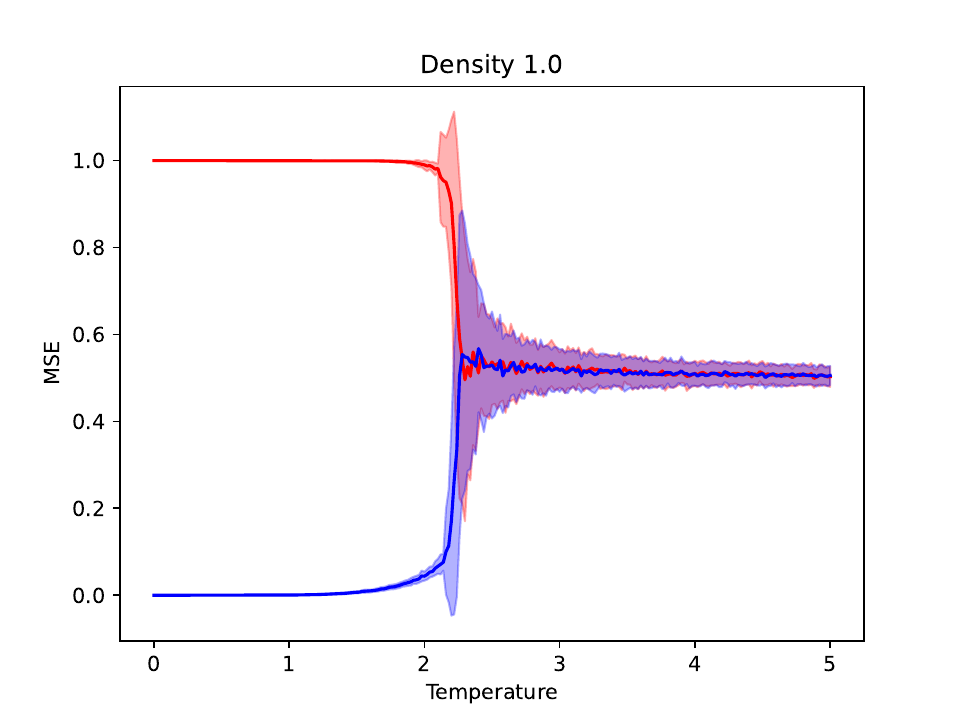}
    \caption{Average MSE corresponding to the reconstruction of Montecarlo configurations after training the autoencoder on a set of synthetic configurations including random patterns.}
    \label{fig:errorSplit_disordered}
\end{figure}

In addition, we observe a strong architectural bias introduced by the activation functions, particularly ReLU, which favor positive activations and therefore implicitly prefer configurations with positive magnetization. This built-in asymmetry leads to a pronounced difference in how the network reconstructs “up” and “down” configurations, despite their physical equivalence. As shown in Fig. \ref{fig:errorSplit_disordered}, at low temperatures, the reconstruction error is minimal for “up” configurations and significantly larger for “down” ones, despite the latter being structurally simpler than high-temperature disordered states.

Strikingly, this effect results in “down” magnetized configurations—nominally more ordered—exhibiting higher reconstruction error than disordered paramagnetic configurations. In this regime, the architecture-induced symmetry bias overrides the complexity advantage, revealing a nontrivial interplay between representational difficulty and model priors.

Although the framework no longer fits the traditional anomaly detection mold, the reconstruction dynamics remain highly informative. They expose the network’s sensitivity not only to phase structure and intrinsic dimension but also to design choices that embed physical asymmetries.

In summary, relaxing the anomaly detection condition provides a broader perspective on how neural networks encode and reconstruct physical phases. The asymmetries in reconstruction error encode valuable signals about both intrinsic data complexity and implicit architectural preferences, offering a robust tool to explore magnetic ordering even in challenging regimes.

\section{Conclusions}
\label{sec:conclusions}

In this work, we presented a systematic exploration of phase classification in the diluted Ising model using neural networks trained exclusively on synthetic data. We focused on three key principles: computational efficiency, synthetic data generation, and physics-informed learning design. The proposed methodology is both lightweight and scalable, providing a low-cost alternative to traditional training pipelines that rely on large experimental or simulated datasets.

Our results show that even simple Dense Neural Networks (DNNs) trained on small catalogs of idealized spin configurations can successfully identify magnetic phases and detect phase transitions. The networks demonstrate strong generalization capabilities, accurately classifying Montecarlo configurations with temperature and dilution values not included in the training set. This finding highlights the utility of synthetic datasets to produce models that extrapolate well beyond their initial training regime.

The second part of our study introduced an unsupervised approach based on convolutional autoencoders (CAEs). By training on ideal configurations representing broken symmetry, the autoencoder effectively distinguished ordered from disordered phases using a reconstruction error signal. This anomaly detection strategy required no labeled data and offered a clear signature of the transition, with sensitivity to both thermal and dilution-induced disorder.

A key innovation of our approach lies in the incorporation of architectural biases to guide learning. In particular, the use of ReLU activation functions introduced an implicit preference for configurations with positive magnetization, magnifying the network’s ability to detect spontaneous $\mathbb{Z}_2$ symmetry breaking. This bias, combined with the exclusion of disordered configurations from training, enabled the model to detect phase boundaries without explicitly encoding physical rules or requiring Montecarlo data for training.

When spin dilution was introduced, the autoencoder—though never trained on configurations with missing spins—continued to detect phase transitions and estimate the percolation threshold. The disappearance of the low temperature reconstruction error splitting provided a robust signal for the loss of magnetic order, consistent with percolation physics.

Finally, by relaxing the strict anomaly detection framework and including both ordered and disordered configurations in the training set, we gained further insight into how reconstruction performance reflects not only intrinsic data complexity but also architectural and symmetry-related biases. While disordered configurations are inherently more complex, their reconstruction error remains intermediate. In contrast, configurations from the “down” sector of the ordered phase—structurally simple but disfavored by the ReLU-based architecture—exhibit the highest reconstruction error. This asymmetry reveals how the network's performance is shaped by an interplay between symmetry breaking, model design, and training data composition. Rather than serving as a strict anomaly detector, the autoencoder functions here as a probe of phase structure, encoding in its reconstruction dynamics both physical and architectural constraints.

Overall, this study demonstrates that efficient training strategies based on synthetic, physics-informed data can yield models that are both interpretable in behavior and powerful in performance. The results open avenues for broader applications of machine learning in physical sciences, especially in scenarios where data is limited, costly, or strategically designed to highlight key symmetries and transitions. Future work may explore the integration of other symmetry groups, alternative generative schemes, or theoretical connections between architectural bias and emergent physical representations.

\section{aknowledgments}

C. A. Lamas and M. Arlego are partially financed by CONICET (PIP 2332) and UNLP (PID X987)

\bibliographystyle{unsrt}
\bibliography{2024_medina}

\begin{thebibliography}{10}

\bibitem{Rem2019}
Benno~S. Rem, Niklas Käming, Matthias Tarnowski, Luca Asteria, Nick Fläschner, Christoph Becker, Klaus Sengstock, and Christof Weitenberg.
\newblock Identifying quantum phase transitions using artificial neural networks on experimental data.
\newblock {\em Nature Physics 2019 15:9}, 15:917--920, 7 2019.

\bibitem{carrasquilla2017machine}
Juan Carrasquilla and Roger~G Melko.
\newblock Machine learning phases of matter.
\newblock {\em Nature Physics}, 13(5):431--434, 2017.

\bibitem{Wang2016}
Lei Wang.
\newblock Discovering phase transitions with unsupervised learning.
\newblock {\em Physical Review B}, 94:195105, 11 2016.

\bibitem{van2017learning}
Evert~PL Van~Nieuwenburg, Ye-Hua Liu, and Sebastian~D Huber.
\newblock Learning phase transitions by confusion.
\newblock {\em Nature Physics}, 13(5):435--439, 2017.

\bibitem{corte2021exploring}
I~Corte, S~Acevedo, M~Arlego, and Carlos~Alberto Lamas.
\newblock Exploring neural network training strategies to determine phase transitions in frustrated magnetic models.
\newblock {\em Computational Materials Science}, 198:110702, 2021.

\bibitem{ponte2017kernel}
Pedro Ponte and Roger~G Melko.
\newblock Kernel methods for interpretable machine learning of order parameters.
\newblock {\em Physical Review B}, 96(20):205146, 2017.

\bibitem{wang2016discovering}
Lei Wang.
\newblock Discovering phase transitions with unsupervised learning.
\newblock {\em Physical Review B}, 94(19):195105, 2016.

\bibitem{wang2017machine}
Ce~Wang and Hui Zhai.
\newblock Machine learning of frustrated classical spin models. i. principal component analysis.
\newblock {\em Physical Review B}, 96(14):144432, 2017.

\bibitem{gomez2024unsupervised}
FA~G{\'o}mez~Albarrac{\'\i}n.
\newblock Unsupervised machine learning for the detection of exotic phases in skyrmion phase diagrams.
\newblock {\em Physical Review B}, 110(21):214415, 2024.

\bibitem{mendes2021unsupervised}
Tiago Mendes-Santos, X~Turkeshi, M~Dalmonte, and Alex Rodriguez.
\newblock Unsupervised learning universal critical behavior via the intrinsic dimension.
\newblock {\em Physical Review X}, 11(1):011040, 2021.

\bibitem{andreas2025}
Djenabou Bayo, Burak {\c{C}}ivitcio{\u{g}}lu, Joseph~J Webb, Andreas Honecker, and Rudolf~A R{\"o}mer.
\newblock Machine learning of phases and structures for model systems in physics.
\newblock {\em Journal of the Physical Society of Japan}, 94(3):031002, 2025.

\bibitem{wetzel2017unsupervised}
Sebastian~J Wetzel.
\newblock Unsupervised learning of phase transitions: From principal component analysis to variational autoencoders.
\newblock {\em Physical Review E}, 96(2):022140, 2017.

\bibitem{ng2023unsupervised}
Kwai-Kong Ng and Min-Fong Yang.
\newblock Unsupervised learning of phase transitions via modified anomaly detection with autoencoders.
\newblock {\em Physical Review B}, 108(21):214428, 2023.

\bibitem{marashli2025identifying}
Mohamad~Ali Marashli, Henry~Lam Ho~Lai, Hamam Mokayed, Fredrik Sandin, Marcus Liwicki, Ho-Kin Tang, and Wing~Chi Yu.
\newblock Identifying quantum phase transitions with minimal prior knowledge by unsupervised learning.
\newblock {\em SciPost Physics Core}, 8(1):029, 2025.

\bibitem{jang2024unsupervised}
Inhyuk Jang and Arun Yethiraj.
\newblock Unsupervised machine learning method for the phase behavior of the constant magnetization ising model in two and three dimensions.
\newblock {\em The Journal of Physical Chemistry B}, 129(1):532--539, 2024.

\bibitem{hu2017discovering}
Wenjian Hu, Rajiv~RP Singh, and Richard~T Scalettar.
\newblock Discovering phases, phase transitions, and crossovers through unsupervised machine learning: A critical examination.
\newblock {\em Physical Review E}, 95(6):062122, 2017.

\bibitem{strubell-etal-2019-energy}
Emma Strubell, Ananya Ganesh, and Andrew McCallum.
\newblock Energy and policy considerations for deep learning in {NLP}.
\newblock In Anna Korhonen, David Traum, and Llu{\'i}s M{\`a}rquez, editors, {\em Proceedings of the 57th Annual Meeting of the Association for Computational Linguistics}, pages 3645--3650, Florence, Italy, July 2019. Association for Computational Linguistics.

\bibitem{Goyal-sinthetic}
Mandeep Goyal and Qusay~H. Mahmoud.
\newblock A systematic review of synthetic data generation techniques using generative ai.
\newblock {\em Electronics}, 13(17), 2024.

\bibitem{Shumailov2024}
Ilia Shumailov, Zakhar Shumaylov, Yiren Zhao, Nicolas Papernot, Ross Anderson, and Yarin Gal.
\newblock Ai models collapse when trained on recursively generated data.
\newblock {\em Nature |}, 631:755, 2024.

\bibitem{Acevedo2022}
Santiago Acevedo, Carlos~A. Lamas, Alejo~Costa Duran, Mauricio~B. Sturla, and Tomás~S. Grigera.
\newblock On the neural network flow of spin configurations.
\newblock {\em Computational Materials Science}, 213, 3 2022.

\bibitem{camastra2016intrinsic}
Francesco Camastra and Antonino Staiano.
\newblock Intrinsic dimension estimation: Advances and open problems.
\newblock {\em Information Sciences}, 328:26--41, 2016.

\bibitem{Pavioni2023}
G.~L.~Garcia Pavioni, M.~Arlego, and C.~A. Lamas.
\newblock Minimalist neural networks training for phase classification in diluted-ising models.
\newblock {\em Computational Materials Science}, 235, 10 2023.

\bibitem{Zhuang-transfer}
Fuzhen Zhuang, Zhiyuan Qi, Keyu Duan, Dongbo Xi, Yongchun Zhu, Hengshu Zhu, Hui Xiong, and Qing He.
\newblock A comprehensive survey on transfer learning.
\newblock {\em Proceedings of the IEEE}, 109(1):43--76, 2021.

\bibitem{karniadakis2021physics}
George~Em Karniadakis, Ioannis~G Kevrekidis, Lu~Lu, Paris Perdikaris, Sifan Wang, and Liu Yang.
\newblock Physics-informed machine learning.
\newblock {\em Nature Reviews Physics}, 3(6):422--440, 2021.

\bibitem{stauffer2018introduction}
Dietrich Stauffer and Ammon Aharony.
\newblock {\em Introduction to percolation theory}.
\newblock Taylor \& Francis, 2018.

\bibitem{Newman2000}
M.~E.J. Newman and R.~M. Ziff.
\newblock Efficient monte carlo algorithm and high-precision results for percolation.
\newblock {\em Physical Review Letters}, 85:4104, 11 2000.

\bibitem{bengio2017deep}
Yoshua Bengio, Ian Goodfellow, and Aaron Courville.
\newblock {\em Deep learning}.
\newblock MIT press Cambridge, MA, USA, 2017.

\bibitem{géron2022hands}
A.~G{\'e}ron.
\newblock {\em Hands-on Machine Learning with {S}cikit-{L}earn, {K}eras, and {T}ensor{F}low: Concepts, Tools, and Techniques to Build Intelligent Systems}.
\newblock O'Reilly, 2022.

\bibitem{chollet2021deep}
F.~Chollet.
\newblock {\em Deep Learning with Python, Second Edition}.
\newblock Manning, 2021.

\bibitem{Acevedo2021}
S.~Acevedo, M.~Arlego, and C.~A. Lamas.
\newblock Phase diagram study of a two-dimensional frustrated antiferromagnet via unsupervised machine learning.
\newblock {\em Physical Review B}, 103, 1 2021.

\bibitem{Zhang1982}
Yufei Zhang, Mingjing Chen, Zhonglue Wen, al, Hao Hu, Henk W~J Blöte, and Youjin Deng.
\newblock Site percolation threshold for honeycomb and square lattices.
\newblock {\em Journal of Physics A: Mathematical and General}, 15:L405, 8 1982.

\bibitem{Newman2001}
M.~E.J. Newman and R.~M. Ziff.
\newblock Fast monte carlo algorithm for site or bond percolation.
\newblock {\em Physical Review E}, 64:016706, 6 2001.

\end{thebibliography}

\appendix

\section{Autoencoder's Reconstructions}

\begin{figure}[ht]
    \centering
    \includegraphics[width=\linewidth]{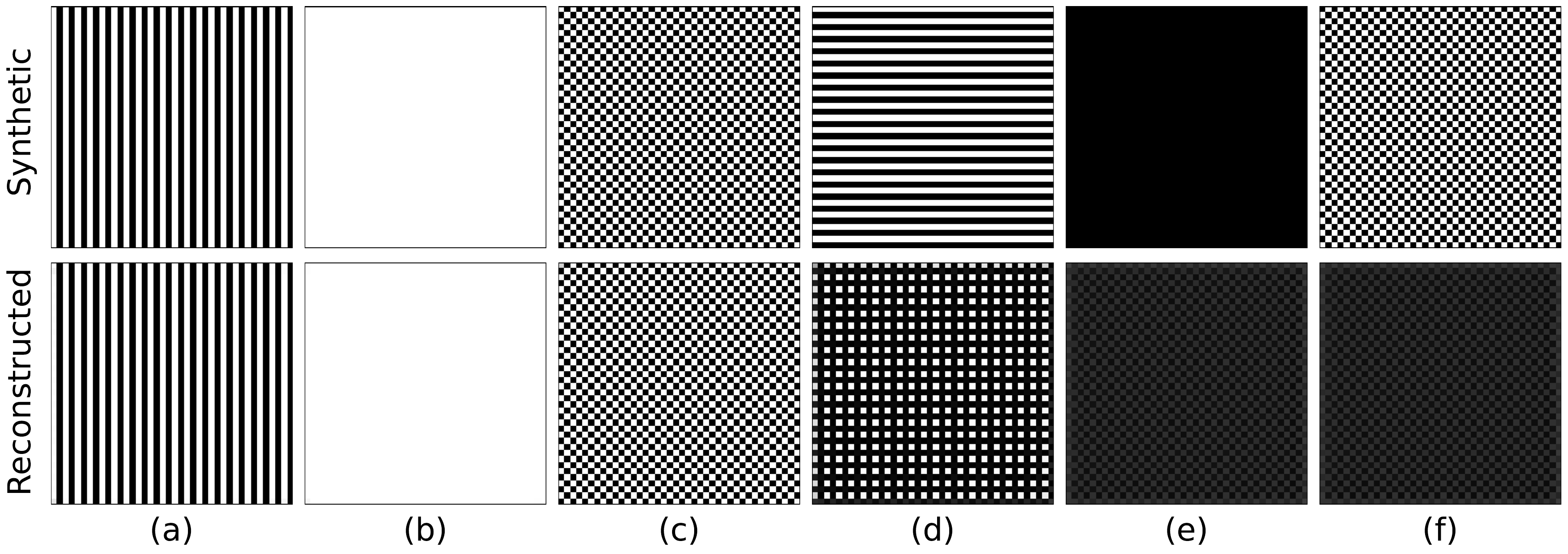}
    \caption{Top:  Synthetic patterns used for training. Bottom: Reconstruction images}
    \label{fig:reconstructed_training}
\end{figure}

In this appendix we briefly discuss the reconstruction of the Montecarlo configurations made by the autoencoder after the synthetic training.
First the autoencoder is trained with synthetic configurations as shown in Fig. \ref{fig:reconstructed_training}. Note how the model favors reconstructions corresponding to positive spin values (represented with white pixels).

\begin{figure}[ht]
    \centering
    \includegraphics[width=\linewidth]{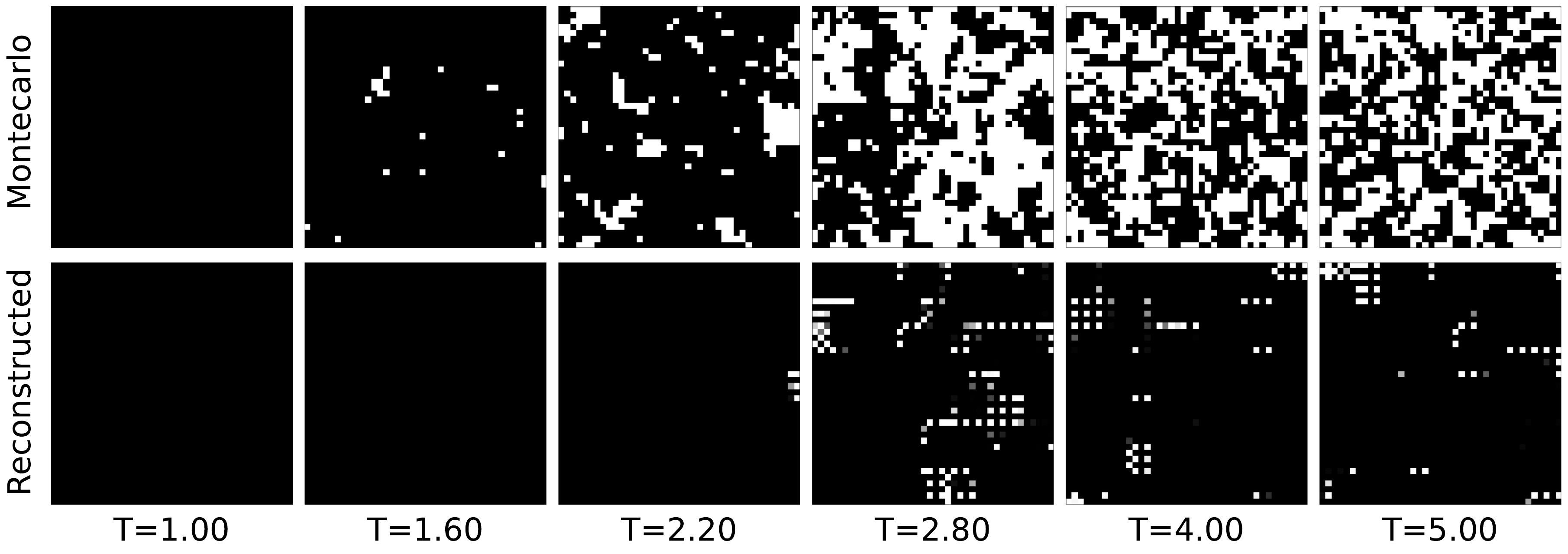}
    \caption{Spin patterns obtained via Montecarlo simulation of the pure Ising Model (upper row) and the autoencoder's reconstruction (lower row). Notice as the reconstruction preserve the overall presence of black and withe spots. }
    \label{fig:reconstructed_montecarlo}
\end{figure}

In Fig.~\ref{fig:reconstructed_montecarlo}, we present representative spin configurations generated via Montecarlo simulations at intermediate temperatures. Although the autoencoder, trained exclusively on perfectly ordered synthetic configurations, cannot reproduce the full magnetic profile of these more complex states, it manages to approximate them by assembling learned patterns into a sort of “collage” that captures the dominant features of the input. As temperature increases, spin configurations become increasingly disordered, and the reconstructions aim to preserve the larger black and white domains characteristic of the underlying structure.

\begin{figure}[ht]
    \centering
    \includegraphics[width=0.9\linewidth]{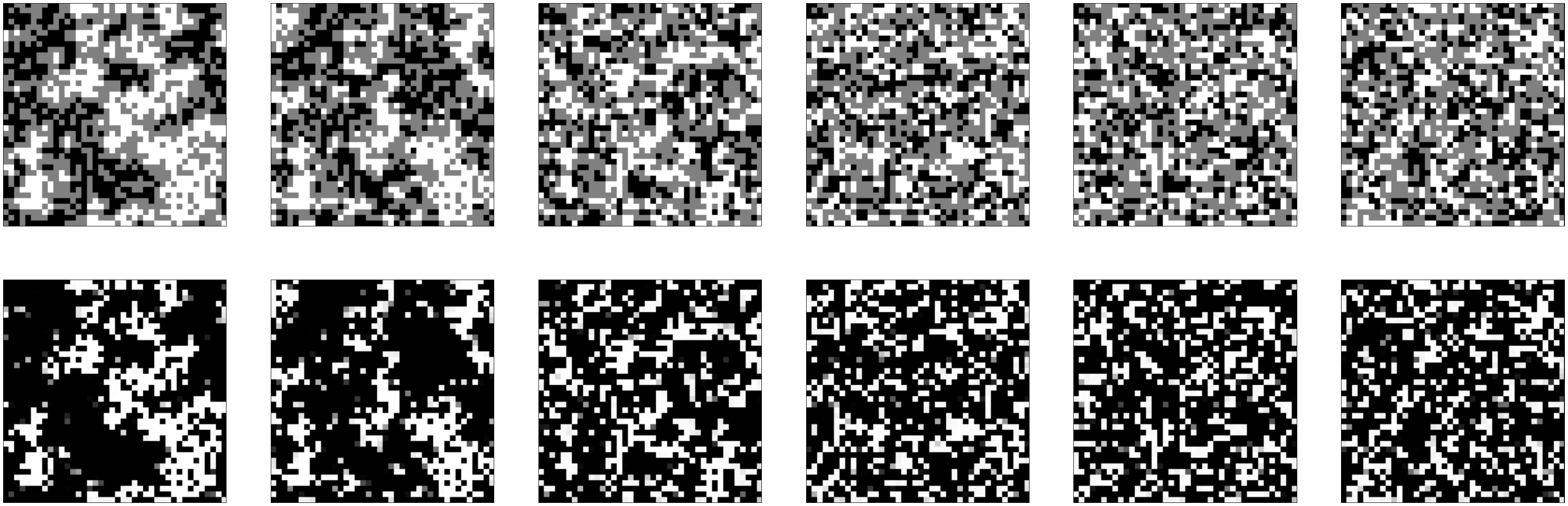}
    \caption{Spin patterns obtained via Montecarlo simulation of the diluted Ising model corresponding to $\rho = 0.55 $ (upper row) and the autoencoder’s reconstruction (lower row).  }
    \label{fig:cae055}
\end{figure}

Figure~\ref{fig:cae055} shows a comparison between input Montecarlo configurations (top row) and their reconstructions (bottom row) for a diluted system.
As expected, the network struggles to accurately reconstruct regions with missing spins, since such patterns were absent from the training data. Nonetheless, the reconstructions exhibit local average values that qualitatively reflect the presence of holes, particularly in regions of low connectivity.

The codes corresponding to the implementation of the neural networks and the data used for their training are available upon request.

\end{document}